\documentclass[preprint,showpacs,preprintnumbers,superscriptaddress,amsmath,amssymb]{revtex4}

\usepackage[dvips]{graphicx}
\usepackage{dcolumn}
\usepackage{bm}

\begin{document}


\title{
Direct observation of localization in the minority-spin-band electrons of magnetite 
below the Verwey temperature
}

\author{Hisao KOBAYASHI}
 \email{kobayash@sci.u-hyogo.ac.jp}
\affiliation{%
Graduate School of Material Science, University of Hyogo, 3-2-1 
Koto Hyogo 678-1297, Japan
}%

\author{Toshihiro NAGAO}%
\affiliation{%
Graduate School of Material Science, University of Hyogo, 3-2-1 
Koto Hyogo 678-1297, Japan
}%

\author{Masayoshi ITOU}
\affiliation{
Japan Synchrotron Radiation Research Institute, SPring-8, Hyogo 679-5198, Japan
}%

\author{Sakae TODO}
\affiliation{
Institute for Solid State Physics, The University of Tokyo, Chiba 277-8581, Japan
}%

\author{Bernardo BARBIELLINI}
\affiliation{
Physics Department, Northeastern University, Boston MA 02115, USA
}%

\author{Peter E. MIJNARENDS}
\affiliation{
Physics Department, Northeastern University, Boston MA 02115, USA
}%
\affiliation{
Department of Radiation, Radionuclides \& Reactors,
Faculty of Applied Sciences,
Delft University of Technology, Delft, The Netherlands
}%

\author{Arun BANSIL}
\affiliation{
Physics Department, Northeastern University, Boston MA 02115, USA
}%

\author{Nobuhiko SAKAI}%
\affiliation{%
Graduate School of Material Science, University of Hyogo, 3-2-1 
Koto Hyogo 678-1297, Japan
}%

\date{\today}

\begin{abstract}

 Two-dimensional spin-uncompensated momentum density distributions, $\rho_{\rm s}^{2D}({\bf p})$s, 
were reconstructed in magnetite at 12K and 300K from several measured directional magnetic Compton profiles.
 Mechanical de-twinning was used to overcome severe twinning in the single crystal sample 
below the Verwey transition.
 The reconstructed $\rho_{\rm s}^{2D}({\bf p})$ in the first Brillouin zone changes from 
being negative at 300 K to positive at 12 K.
 This result provides the first clear evidence that electrons with low momenta in the minority 
spin bands in magnetite are localized below the Verwey transition temperature.
\end{abstract}

\maketitle

\section{\label{sec:level1}Introduction}

 Magnetite, Fe$_3$O$_4$, is the oldest magnetic material recognized 2500 years ago and 
is one of the most fascinating materials in   present-day solid state physics.
 It is because Fe$_3$O$_4$ exhibits many interesting physical properties \cite{walz02,garcia04} 
such as mixed valence and a metal-insulator (MI) transition known as the Verwey transition \cite{Verwey39}.
 Furthermore, Fe$_3$O$_4$ has potential industrial applications in magnetic multilayer devices \cite{kale01} 
since it has full spin polarization with a high magnetic ordering temperature and its electronic structure 
at ambient conditions is predicted by band calculations to be half-metallic\cite{yanase99}.

 Above the MI transition temperature, $T_{\rm v}$, Fe$_3$O$_4$ has an inverse spinel structure 
with cubic symmetry where eight Fe$^{3+}$ ions occupy the A sites with local tetragonal symmetry, 
while the B sites, which locally exhibit octahedral symmetry, are occupied by eight Fe$^{3+}$ 
and Fe$^{2+}$ ions  in a simple ionic chemical formula.
 The results of resonant x-ray scattering \cite{gracia00} show that the Fe ions on the B sites 
are electronically equivalent for time scales lower than 10$^{-16}$ s.
 The MI transition occurs at $T_{\rm v}$ $\sim$ 120 K accompanied by a structural change 
where the conductivity decreases sharply by two orders of magnitude \cite{Verwey39}.
 Early explanations of the MI transition invoked charge ordering on the Fe B sites below $T_{\rm v}$.
 However, NMR relaxation results indicate that the states of Fe ions on the B sites are mixed strongly 
even below $T_{\rm v}$ \cite{novak00}.
 Moreover, recent resonant x-ray and inelastic neutron scattering studies show a fractional charge ordering 
on the Fe B sites below $T_{\rm v}$ \cite{haung06,nazarenko06,schlapp08,mcqueeney06,mcqueeney07}, 
which is predicted by band calculations \cite{leonov04,jeng04}.
 Thus, some studies suggest that only the Fe B sites are involved in the change in valence across the Verwey transition \cite{lodziana07,schlappa08} 
while other investigations claim that the Fe A sites play also an important role \cite{rozenberg06,rozenberg07}.
 Therefore, despite continuing experimental and theoretical efforts in the last 70 years, 
many questions regarding the MI transition still remain open. 

  Magnetic Compton scattering using synchrotron radiation is an established technique for probing 
the spin-uncompensated momentum density distribution in a material with spontaneous magnetization \cite{sakai96,cooper97}.
 The total electron momentum density, $\rho({\bf p})$, is 
\[
\rho({\bf p}) = \rho_{\uparrow} ({\bf p}) + \rho_{\downarrow} ({\bf p}),
\]
where $\rho_{\uparrow} ({\bf p})$ and $\rho_{\downarrow} ({\bf p})$ give the momentum densities of electrons 
in the majority and minority spin bands, respectively, and ${\bf p}$ is the electron momentum.
 Owing to the presence of spin-dependent terms in the scattering cross-section, 
the spin-uncompensated momentum density distribution, 
$\rho_{\rm s} ({\bf p}) = \rho_{\uparrow} ({\bf p}) - \rho_{\downarrow} ({\bf p})$, can be measured 
using circularly polarized x-rays.
 It is obtained in the form of a projection of $\rho_{\rm s} ({\bf p})$ on the x-ray scattering vector, 
the so-called magnetic Compton profile, defined as 
\[
J_{{\rm mag}}(p_z) = \int \int \rho_{\rm s} ({\bf p}) dp_x dp_y. 
\]
 Here the $z$-axis is taken to be parallel to the x-ray scattering vector and $p_z$ is the projection of ${\bf p}$ 
on the $z$-axis.
 The integral of $J_{{\rm mag}}(p_z)$  over $p_z$ yields the spin-magnetic moment of the material.
 For electronic structure studies in a material with spontaneous magnetization, a measurement of 
$\rho_{\rm s} ({\bf p})$ is very useful because $\rho_{\rm s} ({\bf p})$ reflects the spin polarization of 
occupied bands in momentum space. 
 It is possible to image $\rho_{\rm s} ({\bf p})$ from a finite number of $J_{{\rm mag}}(p_z)$s 
using a reconstruction technique.

 In this work we report magnetic Compton scattering studies of a single crystal of magnetite.
 The amplitude of the two-dimensional spin-uncompensated momentum density distribution is shown to develop 
negative excursions above the Verwey temperature.
 We conclude on this basis that a substantial amount of $t_{2g}$ minority electrons become itinerant 
on Fe B sites above $T_{\rm v}$.

\section{\label{sec:level2}Experimental}

 A high-quality single crystal of Fe$_3$O$_4$ was synthesized using the floating zone melting method.
 The crystal was annealed at around 1200 $^{\rm o}$C in the controlled atmosphere of a mixture of CO and CO$_2$ in 
order to homogenize the composition of the specimen and to release internal strain.
 We used a disk-like single crystal sample of approximately 10 mm diameter and 3 mm thickness with its axis 
parallel to the cubic [101] axis.

 Magnetic Compton scattering experiments were carried out at 12 K and 300 K on beamline BL08W at SPring-8, Japan.
 Elliptically polarized x-rays emitted from an elliptical multipole wiggler were monochromatized to 174 keV 
by a bent Si(620) crystal.
 Energy spectra of Compton scattered x-rays from the sample with a scattering angle of 178.0$^{\rm o}$ were measured 
using a 10-segmented Ge solid-state detector with external magnetic field of 20 kOe.
 The estimated momentum resolution is 0.50 $\pm$ 0.01 a.u. full width at half-maximum.

 The directional $J_{{\rm mag}}(p_z)$ is extracted from the difference between two spectra taken 
under the same experimental conditions with alternating 
directions of magnetization of the sample, aligned by an external magnetic field.
 The observed spectra were corrected for the energy-dependent scattering cross section, efficiency of the detector 
and absorption of the sample.

\section{\label{sec:level3}Results}

 Since severe twinning occurs at $T_{\rm v}$ in the single crystalline sample due to the structural change, 
de-twinning of the sample below $T_{\rm v}$ was accomplished by simultaneous application of external 
magnetic field cooling and squeezing to extract directional magnetic profiles $J_{{\rm mag}}(p_z)$ at 12 K.
 An external magnetic field of 20 kOe was applied along the cubic [001] axis to establish the $c$-axis below $T_{\rm v}$.
 The use of a special Cu sample holder enabled us to squeeze the sample       along the cubic [11$\overline{1}$] direction 
as a result of the larger Cu thermal contraction.
 Changes in the extracted $J_{{\rm mag}}(p_z)$ at 12 K were observed below $|p_z|$ $\sim$ 2 a.u. resulting from this mechanical de-twinning. 
 Thus the mechanical de-twinning is effective although we have not estimated the twin ratio in the sample at 12 K 
with and without mechanical de-twinning.
 Figure 1 shows the extracted $J_{{\rm mag}}(p_z)$ along the three cubic principal axes at 12 and 300 K.
 These profiles show a substantial directional dependence of the magnetic spin momentum density in good accord 
with the corresponding theoretical predictions discussed below.

 Figure 2 shows the extracted $J_{{\rm mag}}(p_z)$ at 12 K and 300 K between the cubic [001] and [101] axes.
 Note that the dip in $J_{{\rm mag}}(p_z)$ below $p_z$ $\sim$ 1 a.u. observed from 9$^{\rm o}$ to 27$^{\rm o}$ 
at 300 K disappears at 12 K.
 This result reveals that $\rho_{\rm s}({\bf p})$ changes across $T_{\rm v}$ even though the saturation 
magnetic moment changes by less than 0.1 \% at $T_{\rm v}$ \cite{kakol90}. 
 We have reconstructed the two-dimensional spin-uncompensated momentum density distribution, $\rho_{\rm s}^{2D}({\bf p})$.
 For this purpose, we adopted the so-called direct Fourier-transform method to reconstruct 
$\rho_{\rm s}^{2D}({\bf p})$ from several directional $J_{{\rm mag}}(p_z)$s \cite{suzuki89,tanaka93}.
 In the reconstruction procedure, we assumed the cubic [010] axis to possess a four-fold rotational symmetry, 
even though Fe$_3$O$_4$ at 12 K has a lower than cubic symmetry \cite{iizumi82,wright01}.
 This assumption was made as we measured $J_{{\rm mag}}(p_z)$s only from the cubic [001] to the [101] axis 
due to the mechanical de-twinning using the special Cu sample holder.

 The reconstructed $\rho_{\rm s}^{2D}({\bf p})$s are shown in Figs. 3(a) and (b), which correspond to 
projections of $\rho_{\rm s}({\bf p})$ on the cubic (010) plane.
 The reconstructed $\rho_{\rm s}^{2D}({\bf p}_{(010)})$ looks like a volcano at 12 and 300 K, 
where ${\bf p}_{(010)}$ represents the projection of ${\bf p}$ on the cubic (010) plane.
 At 300 K, the reconstructed $\rho_{\rm s}^{2D}({\bf p}_{(010)})$ in the crater region (below $|p_{[001]}|$ and 
$|p_{[100]}|$ $\sim$ 0.4 a.u.) is negative, while highly positive values are found  
in the region of $|{\bf p}_{(010)}|$ $\sim$ 1.7 a.u. with peaks at $|p_{[001]}|$ = $|p_{[100]}|$ $\sim$ 1.2 a.u., 
where $p_{[001]}$ and $p_{[100]}$ represent the components of ${\bf p}$ along the cubic [001] and [100] axes, respectively.
 The reconstructed $\rho_{\rm s}^{2D}({\bf p}_{(010)})$ below $|p_{[001]}|$ and 
$|p_{[100]}|$ $\sim$ 0.4 a.u. becomes positive at 12 K as seen in Fig. 3 (b).
 Highly positive values in the reconstructed $\rho_{\rm s}^{2D}({\bf p}_{(010)})$ at 12 K appear 
in the region of $|{\bf p}_{(010)}|$ $\sim$ 1.6 a.u. with peaks at $|p_{[001]}|$ = $|p_{[100]}|$ 
$\sim$ 1.1 a.u., similar to the situation at 300 K.

\section{\label{sec:level4}Discussion}

 In a 3$d$ transition metal compound with spontaneous magnetization, the spin-magnetic moment is 
the difference in occupancy between the majority and minority spin bands, which mostly have 3$d$ character.
 Electrons in fully occupied majority and minority spin bands do not contribute to the spin-magnetic moment.
 However, at any point in momentum space spin components of these band electrons are not completely compensated  because hybridization effects depend on the spin state of the band electrons. 
 These uncompensated spin components contribute to $\rho_{\rm s}({\bf p})$ \cite{nagao08}. 
 Accordingly, $\rho_{\rm s}({\bf p})$ is the spin-uncompensated component of all occupied band electrons 
and the area under $J_{{\rm mag}}(p_z)$ is the spin-magnetic moment of the compound.

Fe$_3$O$_4$ is ferrimagnetic below 858 K where the magnetic moments of the Fe A and B sites 
in the inverse spinel structure are antiparallel, resulting in a saturation magnetic moment of 
about 4 $\mu_B$/formula unit, which does not change drastically at $T_{\rm v}$ \cite{kakol90}.
The magnetic electrons are mostly 3d electrons from the Fe sites.
Consequently, the extracted difference in the reconstructed $\rho_{\rm s}^{2D}({\bf p}_{(010)})$ 
results from a change in the band electrons with mainly 3$d$ character.

 Since the symmetry of Fe$_3$O$_4$ above $T_{\rm v}$ is cubic, the negative values below $|p_{[001]}|$ and 
$|p_{[100]}|$ $\sim$ 0.4 a.u. in the reconstructed $\rho_{\rm s}^{2D}({\bf p}_{(010)})$ at 300 K stem from 
electrons in the minority spin bands below $|{\bf p}|$ $\sim$ 0.7 a.u., a region in the first Brillouin zone of 
the inverse spinel structure. 
 Accordingly, these electrons contribute to the electrical conductivity of Fe$_3$O$_4$ because they have itinerant character.
 The reconstructed $\rho_{\rm s}^{2D}({\bf p}_{(010)})$ below $|p_{[001]}|$ and $|p_{[100]}|$ $\sim$ 0.4 a.u. 
is positive at 12 K and the positive values in the reconstructed $\rho_{\rm s}^{2D}({\bf p}_{(010)})$ 
in the region of $|{\bf p}_{(010)}|$ $\sim$ 1.6 a.u at 12 K are smaller than those at 300 K.
 The population of electrons with itinerant character decreases discontinuously at $T_{\rm v}$ reflecting 
the remarkable decrease in the conductivity \cite{Verwey39}.
 Therefore, the observed difference in the reconstructed $\rho_{\rm s}^{2D}({\bf p}_{(010)})$ provides 
clear experimental evidence that electrons with $|{\bf p}|$ $<$ 0.7 a.u. in the minority spin bands 
are localized below $T_{\rm v}$ and that the electronic structure of Fe$_3$O$_4$ with the inverse spinel structure 
is half-metallic in character.
 The MI transition is characterized by the localization of the electrons in the minority spin bands 
which contribute to the spin magnetic moment of Fe$_3$O$_4$.
 This MI transition with the structural change does not strongly affect the electrons with high momenta, 
which have a localized character even at 300 K.
 These results are consistent with a small change in the saturation magnetic moments at $T_{\rm v}$ \cite{kakol90}.
 The behavior of the reconstructed $\rho_{\rm s}^{2D}({\bf p}_{(010)})$ at 12 K cannot be understood using 
the simple ionic model with the inverse or normal spinel structure.

 We have evaluated the number of electrons, $n_{\rm e}$, whose wave functions localize from the positive region 
in the difference of reconstructed $\rho_{\rm s}^{2D}({\bf p}_{(010)})$ between 12 K and 300 K.
 The value of $n_{\rm e}$ is found to be 0.590(2) per unit cell (containing two Fe B sites) and should be compared 
to the charge ordering on the iron B sites of about 0.4 electrons per unit cell \cite{leonov04,jeng04,lodziana07,schlappa08}.
 The higher value of $n_{\rm e}$ indicates that the iron-oxygen hybridization plays also a role in the MI transition.

 We have carried out KKR band structure and momentum density calculations on Fe$_3$O$_4$ in the inverse spinel structure within 
the framework of the local density approximation (LDA) in order to gain insight into the nature of 
$\rho_{\rm s}^{2D}({\bf p})$ \cite{li07}.
 A reasonable level of agreement is seen in Fig. 1 between the computed $J_{{\rm mag}}(p_z)$ and the measured spectra 
at 300 K along the three cubic axes, especially above $p_z$ $\sim$ 2 a.u. 
 Figure 3(c) shows the theoretical $\rho_{\rm s}^{2D}({\bf p})$ after it has been broadened to account 
for experimental resolution.
 The features of the calculated $\rho_{\rm s}^{2D}({\bf p}_{(010)})$ look similar to those of 
the reconstructed $\rho_{\rm s}^{2D}({\bf p}_{(010)})$ at 300 K.
 However, in contrast with the experimental spectrum, the positive peaks at $|p_{[001]}|$ = $|p_{[100]}|$ $\sim$ 1.2 a.u. 
in $\rho_{\rm s}^{2D}({\bf p})$ are sharper in the calculations and the value of the calculated $\rho_{\rm s}^{2D}({\bf p})$ 
at $|{\bf p}_{(010)}|$ $\sim$ 0 is almost zero.
 These discrepancies cannot be understood within the ionic picture, but could arise from exaggerated hybridization 
in the LDA and/or correlation effects missing in our LDA-based calculations.
 As shown in Fig. 4, our computed spectrum before it is convoluted with the experimental resolution does display 
a negative value of the spin momentum density in the first Brillouin zone due to the presence of minority $t_{2g}$ 
itinerant electrons appearing above $T_{\rm v}$, but this negative excursion is smeared out by the effect of 
resolution broadening.
 The observed deeper crater-like feature in the reconstructed $\rho_{\rm s}^{2D}({\bf p}_{(010)})$ could be explained 
by the partial occupation of $d_{3z^2-r^2}$ orbitals reflecting correlation effects beyond the LDA, 
since the majority spin $d_{3z^2-r^2}$ orbitals give a positive contribution to $\rho_{\rm s}^{2D}({\bf p}_{(010)})$ 
at $|{\bf p}_{(010)}|$  $\sim$ 0.
Correlation effects are known to produce partial occupation of the natural orbitals and 
smearing of the 
electron momentum density \cite{bba01}.

\section{\label{sec:level5}Conclusion}

We have carried out magnetic Compton scattering experiments to investigate the spin-uncompensated electron momentum 
density $\rho_{\rm s}({\bf p})$ of Fe$_3$O$_4$ at 12 K and 300 K. 
Mechanical de-twinning was used to overcome severe twinning in the single crystal sample below $T_{\rm v}$.
The reconstructed $\rho_{\rm s}^{2D}({\bf p}_{(010)})$ below $|p_{[001]}|$ and $|p_{[100]}|$ $\sim$ 0.4 a.u. 
changes from being negative at 300 K to positive at 12 K. 
This result reveals that electrons in the minority spin bands with $|{\bf p}|$ $<$ 0.7 a.u. are localized 
below $T_{\rm v}$, while electrons with high momenta, which have a localized character at 300 K, are not affected 
strongly by the MI transition.
Interestingly, the observed behavior of the reconstructed $\rho_{\rm s}^{2D}({\bf p}_{(010)})$ at 12 K 
cannot be understood within the simple ionic model or the standard LDA-based band theory picture, 
and indicates the presence of correlation effects beyond the LDA leading to 
particularly strong smearing of the spin-uncompensated momentum density distribution.

\begin{acknowledgments}

 The experiments were performed at SPring-8 with the approval of the Japan Synchrotron Radiation Research Institute (JASRI) 
under Proposal Numbers 2002B0381-ND3-np and 2003A0308-ND3-np.
 The work at Northeastern University was supported by the US Department of Energy, Office of Science, Basic Energy Sciences 
contract DE-FG02-07ER46352 and benefited from the allocation of time at the NERSC supercomputing center. 
 It was also sponsored by the Stichting Nationale Computer Faciliteiten (NCF) for the use of supercomputer facilities, 
with financial support from the Nederlandse Organisatie voor Wetenschappelijk Onderzoek 
(Netherlands Organization for Scientific Research). 
 H. K. thanks Dr. Y. Tanaka for providing the reconstruction programs.

\end{acknowledgments}

\clearpage
\begin{figure}
\includegraphics{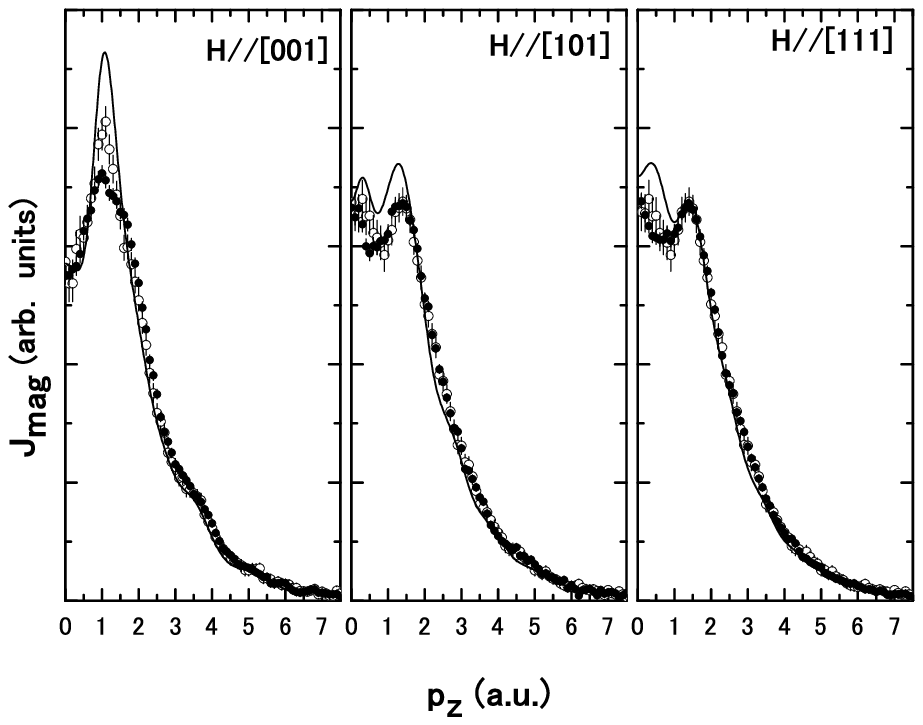}
\caption{\label{fig:mcp_1} 
 $J_{{\rm mag}}(p_z)$ of Fe$_3$O$_4$ along the three cubic principal axes as a function of $p_z$.
Open and filled circles with error bars represent the measured $J_{{\rm mag}}(p_z)$s at 300 and 12 K, 
respectively.
Solid lines give the calculated $J_{{\rm mag}}(p_z)$s in the cubic spinel structure which have been broadened  
to reflect experimental resolution \cite{li07}.
}
\end{figure}

\begin{figure}
\includegraphics{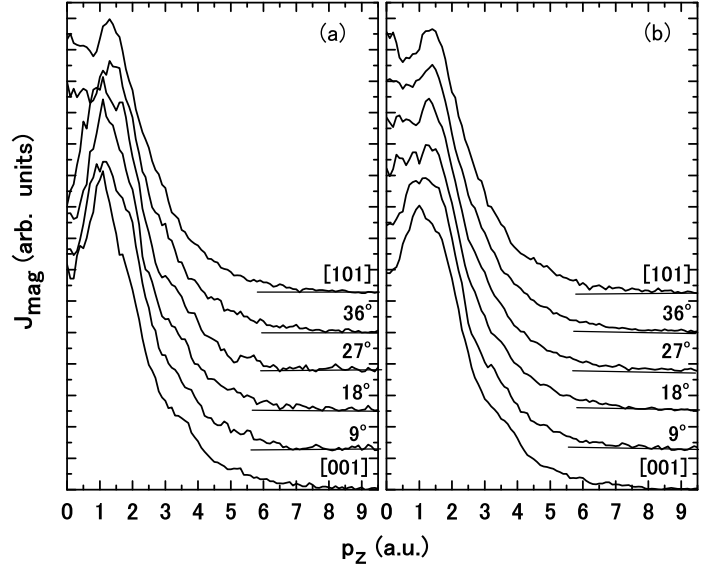}
\caption{\label{fig:mcp_2}
  Measured $J_{{\rm mag}}(p_z)$ of Fe$_3$O$_4$ between the cubic [001] and [101] axes in 
intervals of 9$^{\rm o}$: (a) at 300 K and (b) at 12 K with mechanical detwinning.
}
\end{figure}

\begin{figure}
\includegraphics{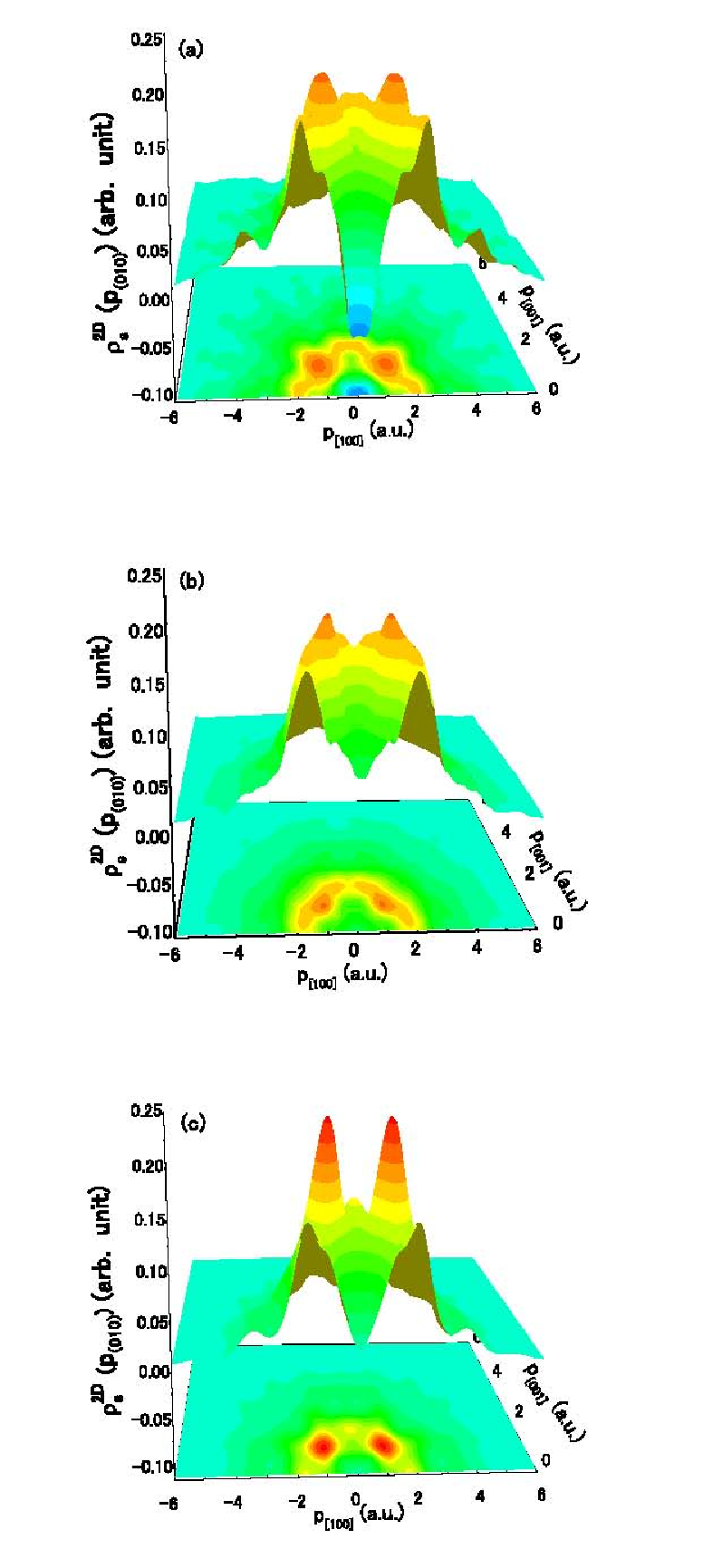}
\caption{\label{fig:rho_s}
(color online) (a) and (b) Reconstructed $\rho_{\rm s}^{2D}({\bf p}_{(010)})$s of Fe$_3$O$_4$ from measured 
$J_{{\rm mag}}(p_z)$s at 300 K and at 12 K, respectively.
  (c) Calculated $\rho_{\rm s}^{2D}({\bf p}_{(010)})$ of Fe$_3$O$_4$ in the inverse spinel structure 
with cubic symmetry which has been broadened to reflect experimental resolution.
}
\end{figure}

\begin{figure}
\includegraphics{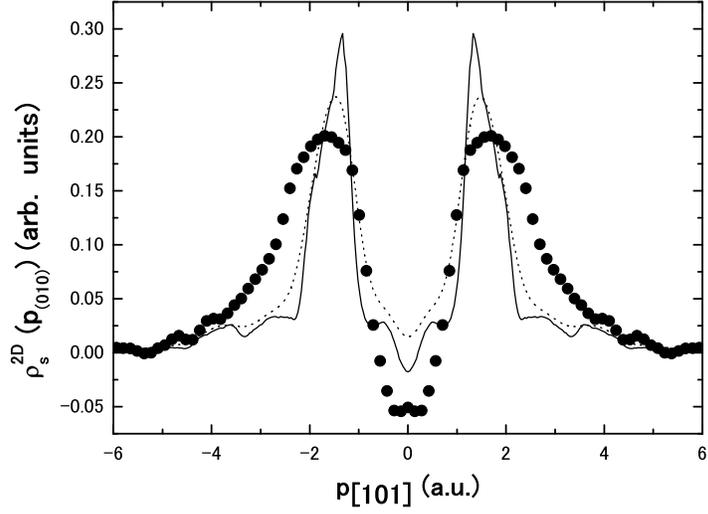}
\caption{\label{fig:rho_101}
 Reconstructed and calculated $\rho_{\rm s}^{2D}({\bf p}_{(010)})$s of Fe$_3$O$_4$ along the cubic [101] axis 
at 300 K and in the cubic spinel structure with and without convolution of experimental resolution, respectively.
 Filled circles represent the reconstructed $\rho_{\rm s}^{2D}({\bf p}_{(010)})$ at 300 K.
 Solid and broken lines give the calculated $\rho_{\rm s}^{2D}({\bf p}_{(010)})$s 
in the cubic spinel structure without and with convolution of experimental resolution, respectively.
}
\end{figure}

\end{document}